# DESTRUCTION AND OBSERVATIONAL SIGNATURES OF SUN-IMPACTING COMETS

John C. Brown
School of Physics and Astronomy, University of Glasgow, Glasgow G12 8QQ
UK (john.brown@glasgow.ac.uk)

Robert W. Carlson
Jet Propulsion Laboratory, California Institute of Technology
Pasadena, CA 91011

Mark P. Toner
41 Castledykes Rd., Castledykes, Dumfries SG1 4SN, UK

**Running Title:** Sun-impacting comets

**Abstract**

Motivated by recent data on comets in the low corona, we discuss destruction of 'sun-impacting' comets in the dense lower atmosphere. Perihelion distances $q \lesssim R_\odot$ and masses $M_o \gg 10^{12}$ g are required to reach such depths.

Extending earlier work on planetary atmosphere impacts to solar conditions, we evaluate the mechanisms and distribution of nucleus mass and energy loss as functions of $M_o$ and $q$, and of parameter $X = 2Q/C_H v_o^2$. $Q$ is the *total* specific energy for ablative mass-loss, $C_H$ the bow-shock heat-transfer efficiency, and $v_o$ the solar escape speed (619 km/s). We discuss factors affecting $Q$ and $C_H$ and conclude that, for solar $v_o$, $X$ is most likely < 1 and solar-impactors mostly ablated before decelerating.

Sun-impacting comets have energies $M_o v_o^2/2 \sim 2 \times 10^{30} \times (M_o/10^{15}$ g) erg, (comparable to magnetic flares $\sim 10^{29-33}$). This is released as a localised explosive airburst within a few scale heights $H \simeq 200$ km of the photosphere, depending weakly on $M_o$, $q$ and $X$. For $X = 10^{-2}$ and $M_o = 10^{15}$ g, a shallow incidence (e.g. polar $\theta \sim \cos^{-1}(0.01)$) Kreutz comet airburst occurs at atmospheric density $n \sim 3\times10^{15}$ cm$^{-3}$ - a height of 700 km (3.5 $H$) above the photosphere (where $n = n_o = 10^{17}$ cm$^{-3}$). The airburst $n$ scales as $\sim (M_o X \cos^3\theta)^{1/2}$ (while height $z$(km) $= 200 \ln(n_o/n)$) so $n$ increases 1000× (700 km deeper) for vertical entry. Such airbursts drive flare-like phenomena including prompt radiation, hot rising plumes and photospheric ripples, the observability and diagnostic value of which we discuss.

1. Introduction

1.1 Background

Over recent decades thousands of (mostly small $M_o \ll 10^{12}$ g) close sun-grazing comets have been observed by white light coronagraphs falling toward the sun, the nuclei of most being fully sublimated by sunlight before or near perihelion. The properties of these small objects, mostly belonging to the Kreutz Group, have been reviewed by e.g. Biesecker et al. (2002), Marsden (2005), and Knight et al. (2010).
In 2011 there was a major breakthrough in observing close sun-grazers when two comets were directly seen 'skimming' the sun's surface and dissipating in the low corona. These observations, made by the AIA UV instrument on SDO (Howard et al. 2008), were of Comets C/2011 N3 (SOHO) and C/2011 W3 (Lovejoy) with perihelia in the low corona ($q = 1.14 R_\odot$ and $1.20 R_\odot$ respectively).

C/2011 N3 (SOHO) crossed the earth-facing disk of the sun and was observed losing mass until the nucleus was fully dissipated over some tens of minutes (i.e. over a distance $\simeq R_\odot$ at speed $v_o$) - see Schrijver et al. (2012) for details of the data and first interpretations. The estimated incident mass was $M_o \lesssim 10^{11}$ g - quite large by Kreutz group standards and large enough to survive sublimation to near perihelion. The much larger Comet Lovejoy 2011 ($M_o \gtrsim 10^{14}$ g - McCauley et al. 2013) passed behind the sun but was observed by SDO in the low corona both on ingress and egress, having survived perihelion passage, though only just

- the emerging material eventually dispersing (Sekanina and Chodas 2012, McCauley et al. 2013).

A third large (~ $10^{14-15}$ g) 'sun-skimmer' - Comet C/2012 S1 ISON was monitored at a wide range of wavelengths including white light, X-Rays and EUV (SDO AIA) as it approached its considerably larger perihelion ($q \simeq 2.5\ R_\odot$) but was undetected by SDO AIA. It seemingly underwent considerable disruption even pre-perihelion, (Sekanina and Kracht 2015) possibly due to insolation-produced outgassing of volatiles and/or by pressure-induced fragmentation, its remnants receding from perihelion in an even more dissipated state than those of C/2011 W3 (Lovejoy). The perihelia of all three of these sun-skimmers lay well inside the Roche tidal limit ($3.4R_\odot$ for the typical mean comet density $\rho_c$ ~ 0.5 g/cm$^3$ - Richardson et al. 2007 - we use here). This could have contributed to their demise though, interestingly, ISON, with the largest $q$, was the most disrupted. The fact that they survived in some form through perihelion might relate to their very high speeds and the resulting short time spent within the Roche limit.

In parallel with these pioneering SDO observations, Brown et al. 2011 (hereafter B11) carried out the first analytic estimations of the different physical regimes of close sun-grazer nucleus destruction as a function of $q$, $M_o$, and of intrinsic nucleus parameters. In particular they showed that an abrupt change occurs in the nucleus destruction mode within the low chromosphere (for comets of large enough $M_o$ and small enough $q$ to reach there) around a certain height above the photosphere which they denoted by $z^*$, the meaning and value of which we discuss in Section 2.4. This is because of the exponential rise of density $n$ with depth over a scale height $H$ of only 200 km. B11 concluded that nuclei of comets with $q > R_\odot + z^*$ (which we term *sun-skimmers*) should lose their mass relatively gradually by insolative sublimation over distances of order $R_\odot$, sometimes reduced by nucleus fragmentation and increased area. Aspects of sun-skimmer destruction have been discussed by Sekanina (2003), B11, Schrijver et al.(2012), Bryans and Pesnell (2012), McCauley et al. (2013), Sekanina and Chodas (2012), Pesnell and Bryans (2014), Sekanina and Kracht (2015).

For comets of higher $M_o$ and smaller $q$ penetrating below height $z^*$, destruction is dominated (as first surmized by Weissman, 1983) by fluid interaction with the dense solar atmosphere there (rather than by insolative sublimation) and we call them *sun-impactors*. This interaction creates a bow shock (Bronshten 1983) with ram-pressure exceeding nucleus material strength and driving it to pancake laterally — see Section 3.2 - accompanied by increasing nucleus ablation and deceleration, and rapid localized deposition of energy and momentum in the atmosphere (Sections 3-6), observational signatures of which are discussed in Section 7. As yet no comets in this sun-impacting regime have been observed in the case of the sun though they should exist — see Section 7.1 — and in any case the sun-impactor theory addressed here can be extended to stellar regimes where they may be common (e.g. Alcock et al. 1986, de Winter et al. 1999, Veras et al. 2014). However, observing sun-impactors is potentially of great interest for a variety of reasons discussed in Section 7 including the possibility of seeing the flash spectrum of an entire comet nucleus as it is vaporized and ionized, and of constraining the value of parameter *X*.

## 1.2 Aims and Structure of this Paper

Our aims in this paper are to improve the theory of sun-impactor destruction, and predictions of their observable signatures, by drawing on and extending the planetary-impact literature (e.g Bronshten's 1983 monograph and extensive subsequent literature) where relevant, and to summarise the physics involved for the solar community who are less familiar with it. Its relevance here stems from the fact that sunlight plays no significant role in destruction of sun-impactors so they resemble planetary atmosphere impactors, though of much higher speed and with different 'target' parameters. It is also hoped to attract active involvement of the planetary impact community in progressing solar impact work.

The structure of the paper is as follows. Section 2 defines and discusses in some detail the values of and uncertainties in the key physical parameters arising in the equations we solve in the sun-impactor problem – the heat transfer and drag coefficients, and 'latent heat' of ablation, and factors affecting their values for the very high solar impact speed $v_o$ compared to planets. Section 2 also amends errors in the values of these parameters adopted by B11. Sections 3, 4, and 5 discuss solutions to the idealized equations of nucleus termination in the deceleration-dominated, ablation-dominated and intermediate regimes. Section 6 reviews the modeling results of sections 3, 4,and 5 while Section 7 addresses some aspects of the observability of comets impacting the Sun and the potential value of such observations.

1.3 Some Points re Jargon, Notation and Regimes

- Most close sun-grazers are members of comet groups — mainly Kreutz Group — resulting from fragmentation of a larger progenitor (possibly primordial) comet in a close encounter with the sun. They thus have $q$ close to $R_\odot$ (cf Section 7.1) so enter the dense solar layers at very shallow angles to the horizontal (thus having small values of $\mu = \cos\theta = (1-q/r)^{1/2}$ for effective impact distance $r(\simeq R_\odot)$ where $\theta$ is the angle to the vertical. Many of our examples are given for such shallow-entry cases but our equations apply to all entry angles including the rarer cases of steep entry and $q$ well inside $R_\odot$.

- Throughout we use the term *vaporize* loosely for conversion of cold dense nucleus material by *any* mechanism into a gas or to liquid droplets.

- In many numerical expressions we utilize a subscript notation denoting parameter values scaled by a suitable power of 10 to typical magnitudes of interest, e.g. $M_o(g) = 10^{15} M_{15}$, $\mu = 10^{-2}\mu_{-2}$, $X = 10^{-2}X_{-2}$. We do so in order that the numerical factor in front of any function of combinations of these parameters gives an immediate indication of its dimensional value for $M_o$ and $\mu$ values in the range we consider of most interest (around $M_{15} \sim 1$, and $\mu_{-2} \sim 1$).

## 2. Nucleus Deceleration and Mass-Loss

### 2.1 Coefficients of Deceleration and Mass Loss

The problem of mass, momentum and energy loss from supersonic bodies impacting a stratified atmosphere is in general complex, and has long been discussed in detail for planetary impacts - see e.g. Bronshten (1983) and references therein, and many subsequent authors. Solving it in depth requires a combination of numerical simulations and judicious analytic approximations in limited regimes. Here and in Section 2.2 we summarize the parameters and equations that let us compare the importance of mass loss versus deceleration. Then in 2.3 we discuss the factors affecting the key parameter values and their consequences for solution regimes under solar impact conditions. Throughout we deal with the near-sun ($r \ll 4\ R_\odot$) regime where the comet speed through the gas is dominated by comet orbital rather than solar wind motion. We also neglect magnetic forces compared to hydrodynamic ones since the low chromosphere is a high-$\beta$ plasma ($nkT \gg B^2/8\pi$). This is *not* the case for sun-skimmers in the low-$\beta$ corona like Lovejoy 2011 whose tail gas motions are strongly striated by magnetic forces (McAuley et al. 2013). We also take the hydrodynamic force acting on an impactor nucleus to be purely along its direction of motion, neglecting lift forces which can become significant for very shallow entry.

If the atmospheric mass density in the atmosphere is $\rho(z)$ (g/cm$^3$) at height $z$ above the photosphere as reference level, the atmospheric momentum and energy flow impinging at comet nucleus speed $v(z)$ on a nucleus of transverse area $A(z) = \pi a^2(z)$ (for a sphere of radius $a$) are $\rho A v^2$ and $\rho A v^3/2$. The fractions of these delivered to the nucleus are respectively the drag coefficient $C_D$ and the heat transfer coefficient $C_H$ (cf. $\Gamma$ and $\Lambda$ in Bronshten's 1983 notation). The incoming supersonic gas is heated and slowed in the shock interaction and the now-subsonic gas is deflected by the nucleus. The thermal and ram pressure of the post-shock gas act on the nucleus and the momentum transfer is large with $C_D$ generally of order unity and commonly assigned a value $C_D = 0.5$ as we will do henceforth. On the other hand $C_H$ is usually << 1 and depends strongly on the quite complex and inefficient processes of energy transfer from shock to nucleus (discussed in Section 2.3). (The incorrect assumption by B11 that $C_H = 1$ would only apply for an extremely tenuous atmosphere with mean free path exceeding the nucleus size). Most of the power $\rho A v^3/2$ incident over area $A$ goes into heating and compressing the decelerated downstream atmospheric flow with only the small fraction $C_H$ available to heat the nucleus and ablate mass from it. Complexities and uncertainties are also features of the second factor determining the nucleus mass loss rate, namely the total specific heat $Q$ (erg/g) required to fully remove unit mass from the nucleus, as discussed in Section 2.3.1.

## 2.2 Deceleration versus Mass-Loss

From the definitions in 2.1, for evolving speed $v$ and area $A \simeq \pi a^2$ for a nucleus with radius $a$, at a point of atmospheric mass density $\rho$, the nucleus mass-loss rate is

$$\frac{dM}{dt} = -\frac{A\rho v^3 C_H}{2Q} \tag{1}$$

while, provided lift forces and the nucleus reaction to the acceleration of ablated matter are small, the momentum equation is

$$M\frac{dv}{dt} = -C_D A\rho v^2 \quad = -\frac{A\rho v^2}{2} \tag{2}$$

Dividing (2) by (1) we can eliminate the $t$ dependence and the $A(t)$, $\rho(t)$ factors and hence obtain, regardless of how $A$ and $\rho$ vary along the path, obtain the equation $d\ln M/dv = vC_H/Q$ for $M(v)$ with solution, for constant $X$ and $M = M_o$ at $v = v_o$,

$$\frac{v(M)}{v_o} = \left[1 + X\ln\left(\frac{M}{M_o}\right)\right]^{\frac{1}{2}} \text{ or } \frac{M}{M_o} = \exp\left[-\frac{1}{X}\left(1 - \left\{\frac{v}{v_o}\right\}^2\right)\right] \tag{3}$$

where the key dimensionless parameter of the problem

$$X = \frac{2Q}{C_H v_o^2} \tag{4}$$

compares the specific heat *required* for ablative mass loss with that $C_H v_o^2/2$ *available* from nucleus kinetic energy via the shock. Smaller $X$ corresponds to higher ablation so our $X$ is *inversely* related by $X = v_o^2/2\sigma$ to Bronshten's (1983, page 14) dimensional "ablation parameter" $\sigma = C_H/2C_D Q$ (g/erg) which *increases* with ablation but, unlike $X$, is *not* normalised relative to the specific incident kinetic energy $v_o^2/2$. The incident speed $v_o$ for the sun is of course much larger than those relevant to planetary impacts and in particular to the meteor physics addressed by Bronshten (1983).

Equation (3) shows that, in the constant $X$ approximation, $v/v_o$ always -> 0 with $M/M_o$ remaining finite viz. $M \to M_{final} = M_o \exp(-1/X)$ but with $M_{final}/M_o$ extremely small for any $X \lesssim 1$ - see Figure 1. Put another way, a fall in mass from $M_o$ to $M_o/e$ by a factor $e$ (exponential), corresponds to a fall in speed from $v_o$ to $v_o(1-X)^{1/2}$. So for $X \ll 1$, fractional deceleration $\Delta v/v_o$ is much less than fractional mass-loss $\Delta M/M_o$ while, for $X > 1$ the residual mass $M_{final}$ exceeds $M_o/e$ when $v \to 0$ and $M_{final}/M_o \to 1$ for $X \gg 1$. (Note that $X \simeq 1$ corresponds to $Q/C_H \simeq v_o^2/2$).

FIG 1

Analytic solution (3) giving the relationship $M(v)$ for constant $X$ can be generalised analytically for special non-constant forms of $X(M,v)$ (such as such as $X(M,v) \sim M^\alpha v^\beta$) allowing analytic integration. However, it does not provide the solutions of coupled Equations (1) and (2) for the time dependence of $M(t)$ or $v(t)$ nor their space dependence, for which we also need to know $A(t)$ and $\rho(t)$. Analytic and numerical aspects of this problem for general $X$ are discussed in Section 5, after we have considered analytically the solutions for $M(t)$, $v(t)$ in the limits of large and of small X in Sections 3 and 4, following our discussion (Section 2.3) of factors affecting the values of $C_H$ and $Q$ (hence $X$) for solar impacts.

The physical interpretation and relationship of the small and large $X$ regimes was first elucidated by Dobrovol'skii (1952) and elaborated by Bronshten (1983), with reference to the fact that the nucleus loses its kinetic energy $KE$ via two routes $dKE/dt = Mvdv/dt + (v^2/2)dM/dt$. The first term here is the slowing, and consequent loss of KE from, the nucleus itself by atmospheric ram pressure at the (*ballistic*) bow-shock, the lost KE going into heating and decelerating the impinging atmosphere. The second term represents the loss of KE of the nucleus in the form of mass ablated from it by heat transfer from the (*ballistic*) shock-front.

The latter energy is transformed and redistributed between the ablated and atmospheric matter via what Bronshten (1983) termed an *ablation shock-front,* the motion of the high density ablated matter acting on the atmosphere like a piston. The behavior of this wave, to which we return in Section 4.2, was viewed by Dobrovol'skii (1952) as like a *blast wave* but by Bronshten (1983, page 123) as more like a *detonation wave*. For the general *X* situation both KE loss processes are always occurring and the interplay between post-shock atmospheric flow, and the flow of ablated nucleus material into it, is complex (Bronshten 1983, pages 122-133). Here we discuss parameters and obtain estimates, for the high-speed solar impact problem, of the distribution of energy deposition with depth, and the consequent dependence of comet termination depths, on nucleus parameters. We do so by considering the limiting cases when one or other KE loss process is dominant.

## 2.3 Values of $Q$, $C_H$ and $X$ for Solar Impacts

### 2.3.1 Value of $Q$

Clearly the minimum possible value of $Q$ is the latent heat $L \approx L_{ice} \sim 2.8 \times 10^{10}$ erg/g needed simply to vaporize nucleus material and it is quite common practice for planetary impacts (e.g. Chyba et al. 1993) to use this value with $C_H$ such that $LdM/dt = C_H A\rho v^3/2$ so that $C_H$ here is the fraction of incident power *reaching the nucleus surface*. $Q \approx L_{ice}$ was also used by B11 for solar impacts. However, it has been emphasized by many (Nemchinov 1967, Bronshten 1983 — page 199, Field and Ferrara 1995, Crawford 1996 and Popova et al. 2000) that $Q \gg L$ is needed to heat and drive mass-loss $dM/dt$ into the flow downstream of the nucleus. An upper limit to $Q$ was suggested by Crawford (1966) to be $Q \sim v_0^2/2$ while Nemchinov (1967) and Popova et al. (2000) characterized $Q$ by the post-shock sound speed. One might also consider the (comparable) post-shock velocity ($v_0/4$ in our case), which would imply $Q = (v_0/4)^2 + L$. The velocity term would dominate $Q$ overwhelmingly for solar $v_o \sim 620$ km/s and in fact dominates for any impact speed $\gg 7$ km/s (e.g. $\sim 60$ km/s for SL-9 Jupiter impacts). This estimate is, however, an upper limit insofar as some of this outflow heating and acceleration may occur when the material is beyond the nucleus radius $a$ used in Equation (1) (cf. Bronshten 1983, pp 92-138).

### 2.3.2 Value of $C_H$

As used in Equation (1) $C_H$ means the fraction of incident atmospheric power available via radiation, convection and conduction from the bow-shock to drive mass *fully* away from the nucleus - i.e. total energy requirement $Q$ erg/g, not just $L$. (In a quasi-steady situation, the mass outflow $dM/dt$ is constant along the flow escaping the nucleus, but the *total* energy per gram delivered from the shock increases from $L$, as material leaves the nucleus surface, to $Q$ as it finally emerges).

The problem of ablation and, in particular, of estimating $C_H$, as a function of the various physical parameters involved and early references to it, were discussed in Bronshten (1983 - especially Chapter 3), particularly in the context of meteors. An extensive literature involving it has developed since, especially in relation to the SL-9 Jupiter impacts (e.g. Chyba et al. 1993, Sekanina 1993, MacLow and Zahnle 1994, Zahnle and MacLow 1994, Field and Ferrara 1995, Svetsov et al. 1995). The ongoing divergence of views on the value of $C_H$ was summarised in Field and Ferrara (1995, pp 958-599) who noted for example that Sekanina (1993) used $C_H$ = 0.6 (almost as large as B11) while Ahrens et al. (1994) proposed $C_H$ = $2 \times 10^{-3}$. Among the many issues complicating determination of the value of parameter $C_H$ are that: (a) Energy transfer from shock to nucleus is limited by the mass loss rate itself through e.g. radiative absorption and blanketing of conductive and convective transport. This makes the heating problem non-linear in incident flux, with $C_H$ a function of that flux; (b) Along the trajectory the density of the target atmosphere, and the speed of the nucleus through it, vary (cf Bronshten, 1983 - page 24), hence so does the incident heat flux; (c) Energy transfer is affected by how the state of the material evolves from solid and fluid through e.g. vapor droplets to ionized gas until it merges with the post-shock atmospheric flow; (d) hydrodynamic instabilities in the ablation flow are likely and can seriously affect the value of $C_H$ (Field and Ferrara 1995). Consequently assigning a single constant parameter value to $C_H$ can only be a first approximation even for a specific set of values for the parameters $M_o$ etc . However, the commonly used constant (mean) $C_H$ approximation is the natural starting point for our initial discussion of the solar impact regime.

Estimates of $C_H$ come from various approaches and with quite diverse results. These include : lab experiments (limited to much smaller masses and speeds than relevant even to large planetary impactors let alone comets impacting the sun); analytic theory and numerical simulations; comparisons of theory with observational data on actual planetary impacts. Many authors argue for very small $C_H$ in modeling large body impacts both with terrestrial planets (e.g. Zahnle 1992) and with Jupiter (e.g. Zahnle and MacLow 1994, Carlson et al. 1995, 1997) while Ceplecha (1966) found $C_H$ = 0.19 from fireball data modeling.

Both lab measurements and theory show that, in the planetary range of parameters, $C_H$ declines with increasing target density $\rho$ and impactor speed $v_o$. Svetsov et al. (1995) discussed and tabulated planetary values for radiative and convective contributions to $C_H$ as functions of atmospheric density $\rho$, and of impactor size $a$ and speed $v_i$. Extrapolating a crude fit $C_H \sim (a\rho)^{-0.5} v_i^{-0.15}$ to their Table to solar photospheric $\rho$, $v_i = v_o$, and to size $a$ = 1 km ($M_o \sim 10^{15}$ g), gives $C_H \sim 4 \times 10^{-4}$. (With $Q = L_{ice} = 2.8 \times 10^{10}$ erg/g this would give a value for $X = 2Q/C_H v_o^2 \simeq 4 \times 10^{-2}$). However, the Svetsov et al. (1995) study was only for relatively low speed (low shock temperature) terrestrial impacts so did not include ionization-related effects. Among these, the thermal conduction contribution to $C_H$ by free hot electrons becomes important at the very high speed $v_o$ and shock temperature $T \sim 4 \times 10^6$ K of sun-impactors. (On the other hand radiative heating should become less important since free-free Planck opacity $\sim T^{-7/2}$). At the solar post-shock temperature $T \sim 4 \times 10^6$ K, the plasma thermal conductivity $\kappa(T) = \kappa_o T^{5/2} \sim 3 \times 10^{10}$ erg/cm/s/K, which exceeds $10^5$ times planetary $\kappa(T)$ even at the post-shock $T$ for Jupiter. The conductive flux is $F_{cond} \sim \kappa(T) T / \lambda$ where the stagnation point stand-off distance $\lambda$ for a sphere of radius $a$ is $\lambda \sim a/4$. For $a \sim 1$ km ($M_o \sim 10^{15}$ g) $F_{cond} \sim 5 \times 10^{12}$ erg/cm$^2$/s, as compared with the incident heating flux is $F_o \sim n\bar{m}v^3/2 \sim 2.6 \times 10^{14}$ $n_{15}$ erg/cm$^2$/s, giving $C_{Hcond} = F_{cond}/F_o \sim 3 \times 10^{-2}$ or about 60 times our extrapolated Svetsov value, with $X$ correspondingly smaller. For a fully pancaked nucleus (see Section 3) the stand-off $\lambda$ value is considerably larger. This will reduce $F_{cond}$ and its absolute contribution to $C_H$, but will also reduce the convective and radiative contributions. Note that with the very short electron mean paths (<< $\lambda$) at these high chromospheric densities, $\kappa(T)T/\lambda$ never approaches the free streaming (saturated heat flux) limit. With the Section 2.3.1 *upper limit* $Q < (v_o^2/2)/16$ the above $C_H$ would imply $X = 2Q/C_H v_o^2 \lesssim 1$ which is in the ablation dominated regime.

A *lower* limit for $X$ itself can be derived from the Field & Ferrara (1995) analysis in the case of the entrainment of droplets formed by heating and turbulent motions at the surface. We write the upper limit condition from their Equation (46) with $C_D = \tfrac{1}{2}$ (Eq. (2)) as

$$\frac{dM}{dt} = \frac{AC_H \varrho v^3/2}{Q} < \tfrac{4}{9}A\varrho v\left[\left(\frac{\gamma-1}{\gamma+1}\right)^{3/2}\left(\frac{\rho_c}{\rho_o(z)}\right)^{1/2}\frac{2\pi H}{30\mu a_o}\right]^{1/2} \qquad (5)$$

which implies $X > X_{min}$ where

$$X_{min} = \tfrac{9}{8}\left[\left(\frac{\gamma+1}{\gamma-1}\right)^{3/2}\left(\frac{\rho_o(z)}{\rho_c}\right)^{1/2}\frac{30\mu a_o}{2\pi H}\right]^{1/2} \simeq 2\times 10^{-3} e^{-\frac{z}{4H}}[\mu_{-2}M_{15}^{\frac{3}{3}}]^{\frac{1}{4}} \qquad (6)$$

There is here some dependence on atmospheric density ~ $\rho^{1/4} = \rho_o^{1/4} e^{-z/4H}$, but in the number density range of interest $n = 10^{14}$–$10^{18}$ $cm^{-3}$ (cf Sections 3–5) this $X_{min}$ varies only by a factor of ~ 10.

## 2.4 The Value of X and Nucleus Destruction Regimes

Although our discussion above suggests that the value of $X$ for solar impacts is likely to be in the ablation dominated regime, given the uncertainties in $C_H$ and $Q$, we henceforth keep $X$ as a free parameter in the range $10^{-4} \lesssim X \lesssim 10$. Clearly more modeling work is needed to improve estimates of $Q$, $C_H$ and $X$ for solar impacts and in Section 7 we discuss whether future observations of sun-impactors might allow estimation of $X$.

As regards the parameter $n^*$, the value of $n$ at height $z = z^*$, this is defined in B11 as the **switch** point from sublimation-dominated to ablation-dominated mass-loss (*sun-skimmer* to *sun-impactor*) occurs. B11 concluded with (incorrect) $Q = L$ and $C_H = 1$ that the **switch occurs at** $n^* \sim 2.5 \times 10^{11}$ cm$^{-3}$ but, for realistic smaller $C_H$ and larger $Q$, this **switch** depth moves down in the atmosphere to $n^* \gtrsim 10^{13-15}$ cm$^{-3}$. This has several important consequences: (i) The paths of the vast majority of close sun-grazers fall entirely in the insolation-dominated (sun-skimmer) regime; (ii) For large masses $M_o$ with $q \lesssim R_\odot$ (sun-impactors), the fractional mass and speed lost in passing through the sublimation domain are small so we can deal with the problem of large sun-impactors essentially as that of injection of the original mass into a dense regime dominated by fluid-interaction mass-loss and deceleration; (iii) Even before ablation exceeds sublimation ($n > n^*$) atmospheric ram-pressure $C_D \rho v^2$ will exceed the estimated strength of cometary nuclei and the nucleus will always be undergoing lateral hydrodynamic expansion (pancaking) during ablation (see Section 3), the increase of $A$ increasing both deceleration and mass-loss rates (Equations (1) and (2)).

In Sections 3-5 respectively we model analytically the two limiting termination regimes (large and small $X$ = deceleration and ablation dominated) and numerically the intermediate regime where both deceleration and ablation effects are comparable. In all three we use the ram-pressure-driven pancaking area solution for $A$ derived in Section 3 and we treat the terminal trajectory as linear (i.e. of constant angle $\theta$ = $\cos^{-1} \mu$ to the vertical). The latter is because almost all of the nucleus-deceleration and mass-loss occur over a radial ($r$) distance of a few scale heights $H$. The fractional change in $\mu = (1-q/r)^{1/2}$ over height $dr = H$ is then $d\mu/\mu = (1/\mu)(d\mu/dr)dr = (1/2\mu)(Hq/r^2) \simeq (H/2R_\odot)/(1-q/r)$ which is small provided $r-q \gg H/2 \simeq 10^{-4} R_\odot$. Thus, so long as the comet end point is many scale heights above perihelion, a linear approximation to the destruction trajectory is good, except in cases of extremely shallow entry when the aerodynamic lift coefficient $C_L$ might become important.

## 3. Analytic Solution in the Deceleration Limit (large $X$)

### 3.1 Basic Solution

The problem of destruction of high mass supersonic bodies in planetary atmospheres has been extensively studied (see e.g. Bronshten 1983 and references in Carlson et al. 1995, 1997 and **B11**), both analytically and numerically, particularly for the impacts of Shoemaker-Levy 9 fragments on Jupiter. A full treatment, especially of the atmospheric response, certainly has to be done numerically (e.g. Chevalier and Sarazin 1994, Carlson et al. 1995) and we are currently undertaking such simulations for the solar case. However, as far as the high $X$ deceleration-dominated regime is concerned, a simple approximate analytic solution exists for an exponential atmosphere of constant density scale height $H$, which happens to be quite a good approximation to the height variation of density in the low chromospheric and photospheric layers of the solar atmosphere where (as we show below) sun-impacting comets with a wide range of $M_o, q$ are destroyed. Geometric height $z$ in the solar atmosphere is conventionally measured from the photosphere (white light optical depth unity) but 'heights' are often also discussed in terms of the particle density $n(z)$ ($cm^{-3}$) (or sometimes overlying particle column density $N(z) = \int_z^{\infty} n(z)dz$). $n(z)$ is related to mass density by $\rho(z)=n(z)\bar{m}$ where $\bar{m}$ is the mean mass per particle. For solar element abundances and in the almost un-ionized low chromosphere/photosphere $\bar{m} \simeq 1.3 m_p$ where $m_p$ is the proton mass. Since the gravity $g$, temperature $T$, and ionization vary much more slowly with $z$ than the density does there, the density distribution is adequately approximated by the hydrostatic isothermal expression

$$\frac{\rho(z)}{\rho_o} = \frac{n(z)}{n_o} = \frac{N(z)}{N_o} = e^{-z/H} \qquad (7)$$

where the scale height $H = \frac{kT}{\bar{m}g} \approx 200$ km is constant and the photospheric reference density is $\rho_o$ = 2.2x10$^{-7}$ g cm$^{-3}$ ($n_o$ = 10$^{17}$cm$^{-3}$).

The analytic solution in the large $X$ limit for the evolution of a supersonic mass impacting such a structure at zenith angle $cos^{-1}\mu$ has been arrived at via several different arguments by a number of authors – Petrov and Stulov (1975), Revelle (1979), Zahnle (1992), Chyba et al.(1993), MacLow and Zahnle (1994), Zahnle and MacLow (1994), Field and Ferrara (1995). It is valid past the point in the atmosphere where ram pressure exceeds the nucleus strength. Except for very large objects the timescale $a/c_s$ < 0.3 $a$(km) s for pressure waves (speed $c_s$) to traverse the nucleus is much less than the timescale $\simeq$ $H/v_o\mu$ $\simeq$ $30/\mu_2$ s on which the ram-pressure grows. Thus the ram-pressure compresses the nucleus longitudinally while the much smaller atmospheric pressure allows free transverse expansion of the nucleus that consequently pancakes at almost constant density. The analytic solution for the radius $a(z)$ of the pancake is then, for $a(z) \gg a_o$

$$a(z) = \frac{H}{\mu}\left[\frac{2\rho(z)}{\rho_c}\right]^{\frac{1}{2}} \tag{8}$$

where in the [ ] expression we have adopted the factor 2 from MacLow and Zahnle (1994), Zahnle and MacLow (1994), rather than the factor 3 in Zahnle's (1992) earlier work, and for the comet nucleus density we will use henceforth $\rho_c = 0.5$ g cm$^{-3}$. (Note that expression (8) for the pancaked disk radius $a$ is independent of $M$ but the disk thickness $= M/\pi\rho a^2$) is $M$ dependent). The nucleus speed is then found by integrating $M_o dv/dt = -\pi a^2(z)C_D\rho v^2$ to be

$$v(z) = v_o \exp(-2Ke^{-\frac{2z}{H}}) \tag{9}$$

where the dimensionless constant

$$K = \frac{\pi C_D \rho_c H^3}{2M_o\mu^3}\frac{\rho_o^2}{\rho_c^2} = \frac{\pi \rho_c H^3}{4M_o\mu^3}\frac{\rho_o^2}{\rho_c^2} \approx \frac{2.4}{M_{15}\mu_{-2}^3} \tag{10}$$

the second algebraic expression and the numerical expression being for $C_D = 1/2$ and with the scaled units discussed in Section 1

$$M_{15} = \frac{M_o}{10^{15}}g \quad \text{and} \quad \mu_{-2} = \frac{\mu}{10^{-2}} \tag{11}$$

chosen such that $M_{15}$ and $\mu_{-2}$ are around unity for largish Kreutz group members.

Approximation (8) is unphysical at small depths (large $z$), implying that $a \to 0$ as $\rho \to 0$ whereas in fact $a \to a_o$ there, so underestimating $a$ for small $\rho$. The heuristic form

$$a(z) = \left[a_o^2 + 2\frac{\rho(z)}{\rho_c}\frac{H^2}{\mu^2}\right]^{\frac{1}{2}} \tag{12}$$

behaves correctly for small and large $\rho$ and is convenient in allowing closed form approximations to integrals involving $A(z) = \pi a^2(z)$ such as Equation (24) below.

3.2 Deceleration-Dominated Distribution of Energy and Momentum Deposition

As the nucleus decelerates, its kinetic energy $E = Mv^2/2$ and momentum $P = Mv$ decline with depth and, by (9), are deposited into the atmosphere via the bow shock with vertical gradients (energy and momentum per unit $z$)

$$E'_{dec} = \left(\frac{dE}{dz}\right)_{dec} = 8KE'_o \exp\left[-2\left(\frac{z}{H} + 2Ke^{-\frac{2z}{H}}\right)\right] \tag{13}$$

and

$$P'_{dec} = \left(\frac{dP}{dz}\right)_{dec} = 4KP'_o \exp\left[-2\left(\frac{z}{H} + Ke^{-\frac{z}{H}}\right)\right] \tag{14}$$

where

$$E'_o = \frac{M_o v_o^2}{2H}; \quad P'_o = \frac{M_o v_o}{H} \tag{15}$$

Setting $dE'/dz = 0$ gives the height and density where $E'$ peaks

$$z_{decMaxE'} = \frac{H}{2}\ln(4K) \approx 100km \times \ln\left(\frac{9.5}{M_{15}\mu_{-2}^3}\right) \tag{16}$$

and

$$n_{decMaxE'} = \frac{n_o}{(4K)^{1/2}} \approx 3.1\times 10^{16} cm^{-3}(M_{15}\mu_{-2}^3)^{\frac{1}{2}} \tag{17}$$

which is shown in Figure 3 together with the mass-loss-dominated results we drive in Section 4.2. The peak energy deposition gradient (per unit $z$) is

$$E'_{max} = \frac{2}{e}E'_o \tag{18}$$

and occurs when the nucleus speed is, by (9)

$$v_{decMaxE'} = \frac{v_o}{e^{1/2}} \tag{19}$$

Also of interest are the values of the disk's lateral radius *a* and its lateral expansion **speed** *u* at the peak *E'* point. Inserting expression (17) for the end-point *n* into approximation (8) for *a(n)* (*a* >> $a_o$) gives

$$\frac{a_{decMaxE'}}{a_o} \approx \frac{20}{\left(\mu_{-2}^3 M_{15}\right)^{1/12}} \tag{20}$$

shown in Figure 2 for a wide range of $\mu_{-2}^3 M_{15}$ to which it is very insensitive. (Equation (20) predicts $a_{decMaxE'}/a_o$ < 1 for very large $M_o$, which is meaningless. However, pancaking solution (20) breaks down at very large $M_o$ because of the pressure wave travel time argument given in Section 3.1).

---
FIGURE 2

---

The disk transverse expansion speed *u(z)* is, using Equations (8),(10),(16),(19)

$$u(z) = \frac{da}{dt} = \mu v \frac{da}{dz} = \left[\frac{2\rho(z)}{\rho_c}\right]^{\frac{1}{2}} \frac{v(z)}{2} \tag{21}$$

which, at $z_{decMaxE'}$, is

$$u_{decMaxE'} = v_o \left[\frac{\mu^3 M_o}{4\pi e^2 \rho_c H^3}\right]^{\frac{1}{4}} \approx 8.7 \times 10^3 (cm/s)(\mu_{-2}^3 M_{15})^{\frac{1}{4}} \tag{22}$$

also shown in Figure 2. This very low *u* value shows that the pancaking is *not* "explosive", involving negligible kinetic energy compared with *v*, but greatly affecting the solution by increasing *A* and hence both the deceleration and ablation rates.

The analysis and results of this Section predict the depth distribution of energy deposition (including the height of the termination point where this deposition peaks) for the limiting case of no ablation with heating exclusively via passage of the atmosphere through the shock. In addition, the solution provides an estimate of the evolving size of the pancaked nucleus which we also use in Section 4. We would emphasise, however, that this pancaking description with a unique pancake radius *a* and lateral expansion speed *u* is idealized in that pancaking is likely to be accompanied by fragmentation (e.g. Field and Ferrara 1995). In addition expressions (14)-(17) for the energy deposition height profile etc are based on the assumption that Equation (7) holds during the energy deposition whereas the intensely heated atmosphere will increasingly undergo expansion as the event proceeds. Numerical treatment will clearly be necessary to follow the time evolution properly.

## 4. Analytic Solution in the Ablation Limit ($X \ll 1$)

### 4.1 Basic Solution

In the absence of significant ram-pressure, ablation would cause the nucleus radius $a(z)$ to shrink, the ablation rate to slow and make the total ablation path very long. However, since (Section 2.4) ram-pressure sets in early, pancaking of the nucleus shortens the ablation path. Although the pancake solution (8) for $a(z)$ is based on a constant mass approximation, its form is mass independent so, at least as far as nucleus geometry is concerned, it can be applied to estimate the ablative mass-loss rate in the presence of pancaking.

The evolution of nucleus mass with depth is best described using the vertical atmospheric column density

$$N(z) = \int_z^\infty n(z')dz' = n(z)H \qquad (23)$$

Using heuristic form (12) for a(z), we find that, in the small X (ablation-dominated) limit, the mass loss gradient with respect to N is

$$\frac{dM}{dN} = -\frac{\pi \bar{m}}{\mu X}\left[a_o^2 + \frac{2\bar{m}HN}{\rho_c \mu^2}\right] \quad (24)$$

with solution

$$M(N) = M_o - \frac{\pi \bar{m}}{\mu X}\left[a_o^2 N + \frac{\bar{m}HN^2}{\rho_c \mu^2}\right] \quad (25)$$

The point $N = N_{ablMaxE'}(M_o)$ at which $M \to 0$ is the solution of the full quadratic Equation (25) for the end depth $N_{ablMaxE'}$, and is also the N of maximum E' (see Section 4.2). One finds that, in the region of interest (maximum dM/dN), the terms arising from the $a_o$ are small and we can approximate the quadratic solution as

$$N_{ablMaxE'} = Hn_{ablMaxE'} = \left(\frac{\mu^3 M_o \rho_c X}{\pi H \bar{m}^2}\right)^{\frac{1}{2}} \quad (26)$$

## 4.2 Mass-loss-dominated Energy and Momentum Deposition Gradients

Even when the mass ablation rate is so high that there is little deceleration of the nucleus itself, the bow-shock still exists (being supersonically decoupled from what happens behind it) and heats the atmospheric gas passing through it, just as in the deceleration-dominated case (Section 3.2). However, for small X, the ablative mass-loss from the nucleus is **deposited and decelerated in** the hot post-shock gas, heating both components (Dobrovol'skii 1952, Bronshten 1983)- cf Section 2.2 — in proportions depending on X. Here we are interested in the distribution of total energy and momentum deposition per unit depth from the ablated mass as it mingles with the post-shock atmosphere, when (for small X) this is the primary energy deposition process. For constant nucleus speed $v = v_o$, the energy *loss* per unit *z from* the nucleus into ablated mass is just $E' = M'v_o^2/2$ (and similarly $M'v_o$ for momentum loss gradient). Then, provided the 'blending' of the ablated and atmospheric gases is rapid enough, the gradient of energy *deposition* by the ablated material *in* the combined flow is just $E'_{abl} = dE/dz = -(v_o^2/2)dM/dz = -(v_o^2/2)n(z)dM/dN$.

For very low ablation rate (large $X$) the momentum and energy exchange would be on the scale of the very small mean free path of a single ablated atom/ion in the shocked gas, so very fast compared to the nucleus destruction time ($\sim H/\mu v_o$). For smaller $X$, however, the ablated mass flow is initially much denser than the post-shock atmospheric flow and will in practice only be decelerated by and blend with the latter once the 'piston' action of this 'ablation-shock' front has been dissipated as the ablation flow density falls to near that of the atmospheric flow (cf. Dobrov'skii 1952, Bronshten 1983, and Section 2.2). For nucleus radius $a$ and atmospheric density $n$ the mass ablation rate is $dM/dt = \pi a^2 C_H n v_o^3$ and, for ablation flow speed $v_1$, at distance $b_1$, the ablation flow number density (nucleons/cm$^3$) $n_1 = (dM/dt)/(4\pi \bar{m} v_1 b_1^2)$ so that $n_1/n = (v_o/Xv_1)(a/2b_1)^2$. (To avoid any ambiguity, note that $v_1$ is the outflow speed from the nucleus of the heated ablating GAS and $n_1$ its density at distance $b_1$ from the nucleus. These are quite distinct from the slow ram-pressure-driven pancaking expansion speed $u = da/dt$ and radius $a$ (Equations (21) and (8)) of the cold dense nucleus itself). Then, for $v_1 > 10$ km/s and $X > 10^{-2}$, expansion of the ablation flow will roughly equalize the densities and stall the ablation front within a distance $b_1 \sim 80a$ or a time $< b_1/v_1 \sim 8$s — very much longer than $d/v_1$ but still less than the nucleus dissipation time ~ $H/\mu v_o$. Consequently the height profile of the energy deposition gradient by the ablation front will still roughly match the profile of the ablative energy loss profile $dE/dz$ from the nucleus.

It is obvious from Equation (24) that the ablative $dM/dN$ increases monotonically with $N$ (and $n(N) = N/H$) up to a peak at $N = N_{ablMaxE'}$. Hence, for near constant $v = v_o$, the energy loss gradient rises monotonically and exponentially with geometric depth on a spatial scale of order $H$, to a sharp peak at $N = N_{ablMaxE'}$, where it terminates, namely at density (with $X=10^{-2}X_{-2}$)

$$n_{ablMaxE'} = \frac{N_{ablMaxE'}}{H} = \left(\frac{\mu^3 M_o \rho_c X}{\pi H^3 \bar{m}^2}\right)^{\frac{1}{2}} = 2.0 \times 10^{15} (\mu_{-2}^3 M_{15} X_{-2})^{1/2} \quad (27)$$

FIG 3

and equivalent height $z_{ablMaxE'} = H \ln(n_o/n_{ablMaxE'})$. Expression (27) for $n_{ablMaxE'}$ is plotted in Figure 3 versus $\mu_{-2}^3 M_{15}$, for $X = 0.001, 0.01, 0.1, 1.0$ along with the previous results of Equation (17) for $n_{decMaxE'}$ in the deceleration-dominated case. Also shown on the Figure are the heights $z$ corresponding to the $n$ values.

Inserting $n = n_{ablMaxE'}$ in

$$E'_{abl} = \frac{M'_{abl} v_o^2}{2} = \frac{\pi (\bar{m} v_o H n)^2}{X \rho_c \mu^3} \quad (28)$$

gives the maximum $E'_{abl}$ value as

$$E'_{ablMax} = \frac{M_o v_o^2}{H} \approx 2 \times 10^{24} M_{15} \text{ erg/cm} \quad (29)$$

Equations (27) and (29) for this low $X$ regime can be compared with (17) and (18) for the large $X$ regime.

It is important to note that that the quantities $E'_{abl}$ here, and $E'_{dec}$ in Section 3, are energies deposited per unit *vertical* distance *z*. The total energies deposited per unit length *along* the nucleus trajectory are $\mu E'_{abl}$ and $\mu E'_{dec}$. To obtain the heat $\mathcal{E}$ delivered per unit *volume* we need to divide $\mu E'_{abl}$ or $\mu E'_{dec}$ by the cross-sectional area *A* of that volume. For example by Equation (29) for $E'_{ablmax}$, the peak energy deposited per unit length *along* the nucleus trajectory is $\sim \mu M_o v_o^2/H$ and per unit volume is $\mathcal{E} = \mu M_o v_o^2/HA \sim 4 \times 10^{11} \mu_{-2} M_{15}/A(km^2)$ erg/cm$^3$. If one evaluates $A = \pi a^2$ for *a* at the $E'_{max}$ position using Equation (20) then, near the nucleus of a $10^{15}$ g object, $A(km^2) \sim 1000$ and $\mathcal{E} \sim 4 \times 10^8$ erg/cm$^3$ while for very large magnetic flares the typical energy release is $\lesssim 10^5$ erg/cm$^3$ (Tandberg-Hanssen & Emslie, 1988). This high $\mathcal{E}$ is because of the very small dissipation volume in the comet case. On the other hand the peak dissipation sites of comet impacts are typically much deeper and of much higher densities than those of flares (Tandberg-Hanssen & Emslie, 1988). We discuss the consequences of this for observational signatures of comet impacts in Section 7 (specifically 7.3.2). The main result of this Section is to evaluate how the distribution of energy deposition with height (atmospheric *n* value), and in particular the height (*n* value) of peak deposition, change relative to the pure deceleration case when ablation becomes increasingly dominant with declining $X$ ($\lesssim 1$).

## 5. Heuristically Blended and Numerical Results including both Deceleration and Ablation

### 5.1 Heuristically Blended Solution

One can obtain an estimate of the termination depth $n_{end}$ when both deceleration and mass-loss are included by using the heuristic parametric reciprocal mean of the limiting deceleration only and mass-loss only results (Equations (17) and (27)) viz.

$$n_{end}(X) = \left[ 1/n_{ablmaxE'}(X) + 1/n_{decMaxE'} \right]^{-1} = 3.1 \times 10^{16} (\mu_{-2}^3 M_{15})^{1/2} \left[ 1 + 1.5/X^{1/2} \right]^{-1} \quad (30)$$

shown in Figure 4 with a dashed line to indicate the lower limit value of $X$ based on the analysis by Field and Ferrara (1995) (Section 2.3.2).

Fig 4

### 5.2 Numerical Solution

The actual solution of Equations (1) and (2) for $M$, $v$ as functions of depth for any constant $X$ and specified $M_o$, $q$ can be obtained by numerical integration and we can in fact assimilate arbitrary $M_o$, $q$ into scaled variables as follows. A similar treatment can be applied for *variable X = X(M,v,N)* (i.e. $X = X(\zeta, \eta, \xi)$) if one has physical information on $X(M,v,N)$ or wishes to explore the effect of variable $X$ using heuristic parametric forms.

We can re-write the Equations (2) and (3) (with $dN = \mu n v dt = \mu \rho v dt / \overline{m}$) as

$$M(N) \frac{dv}{dN} = -\frac{A \overline{m} v}{2\mu} \quad (31)$$

and

$$\frac{M(N)}{M_o} = exp\left[ -\frac{1 - v^2(N)/v_o^2}{X} \right] \quad (32)$$

29which, on eliminating M, give

$$\frac{dv}{dN} = -A(N)\frac{\bar{m}\,v_o}{2M_o\mu}\frac{v}{v_o}\exp\left[\frac{1-v^2(N)/v_o^2}{X}\right] \tag{33}$$

$A(N)$ is adequately given by Equation (8) as

$$A(N) = \frac{\pi\left(\frac{H}{\mu}\right)^2 N\bar{m}}{2H\rho_c} \tag{34}$$

Differential Equation (33) with (34) can be integrated numerically to get $v(N)$ with $v(0) = v_o$ and hence $M(N)$ by (32) for any values for the parameters $X$ and $\mu^3 M_o$. In fact the parameter $\mu^3 M_o$ can be further assimilated into a scaled dimensionless depth variable

$$\xi(N) = \frac{\pi\bar{m}^2 H N^2}{2\rho_c M_o} = \frac{0.34}{\mu_{-2}^3 M_{15}}\left[\frac{N}{10^{24}\,cm^{-2}}\right]^2 \tag{35}$$

Along with dimensionless variables $\eta(\xi) = v(N)/v_o$ for velocity and $\zeta(\xi) = 1-M(N)/M_o$ for fractional mass-loss the equations become

$$\frac{d\eta}{d\xi} = -\eta(\xi)\exp\left[\frac{1-\eta^2(\xi)}{X}\right] \tag{36}$$

$$\zeta(\xi) = 1 - \exp\left[-\frac{1-\eta^2(\xi)}{X}\right] \tag{37}$$

*involving only the single parameter X.* (Note that for $X \gg 1$ Equation (36) has analytic solution $\eta = e^{-\xi}$ in agreement with deceleration-dominated solution Equation (9)).

Figure 5

In Figure 5 we show the numerical solutions $\eta(\xi)$ and $\zeta(\xi)$ for values of $X$ = 0.01, 0.1, 1 and 10. These illustrate *the* change-over from mass-loss-dominated to deceleration-dominated as $X$ increases. In short, for $X \lesssim 1$ the mass-loss fraction approaches unity ($\zeta=1$) before significant deceleration with very little mass left when deceleration is complete, while for $X \gtrsim 1$ deceleration is high enough that $v \to 0$ with a substantial residual stationary mass $(1-\zeta)M_o$.

## 6. Discussion and Conclusions re Modelling

We have discussed the relative importance of mass-loss and deceleration in terminating the lives of 'sun-impacting' cometary nuclei, i.e. those of large enough incident mass $M_o$ and small enough perihelion $q$ to reach depths where ram-pressure causes pancaking, and mass-loss by ablation exceeds that by insolation. The *relative* importance of mass-loss and deceleration is independent of pancaking since each scales as area $A$, so depends solely on the single parameter $X = 2Q/C_H v_o^2$. Here $Q$ is the total heat energy required to vaporize *and fully remove* unit mass from the nucleus and $C_H$ is the fraction of the incoming power density available to provide that total energy $Q$. For any $X \lesssim 1$, ablative mass-loss dominates and is almost complete before much deceleration, while for large $X$ deceleration is complete before significant mass-loss.

For realistically small values of $C_H$ and for planetary impact speeds, $X$ is quite large and deceleration is dominant even for the small values of $Q \sim L_{ice}$ commonly used. In the solar case, however, $v_o^2$ is > 100 times larger, so that $X$ is expected to be small and mass-loss to be dominant though we have treated both cases since $X$ is still rather uncertain.

We have shown that for all cases the nucleus termination depth can be described approximately by the heuristic blending expression (30) (Figure 4) with an upper limit approached for large $X$ (deceleration-dominated). For any given value of $\mu_{-2}M_{15}^{1/3}$, Equation (30) shows the atmospheric density at the termination depth to vary only by a factor of 10 even for a factor of 100 variation in *X* and to be within a few scale heights above the photosphere for $0.01 < X < 1$ in a typical case where $\mu_{-2} M_{15}^{1/3} = 1$. Vertical entry ($q=0$, $\mu=1$) would increase the end point density by a factor of 1000 (about 7 scale heights ~ 1400 km ~ 2 arcsec geometrically lower) while increasing $M_o$ from $10^{15}$ g (~ Lovejoy mass) to $10^{18}$ g (~ Hale-Bopp mass) would increase it by a factor of 100.

In absolute terms, for an ablation-dominated case with $X = 0.01$ and entry at $\mu = 0.01$ ($q/R_\odot \simeq 0.9999$), the minimum incident mass reaching the photosphere is around $10^{18}$ g, approaching that of Hale-Bopp, while for vertical entry it is about $10^{15}$ g (~ mass of Lovejoy 2011). A major distinguishing feature of the ablation dominated $X < 1$ regime is the high metallicity of the hot plasmas created, as discussed in Section 7.

Finally, an interesting feature of all of our analysis is the almost universal occurrence of the parameters $M_o$ and $\mu$ in the combination $M_o^{1/3}\mu = M_o^{1/3}\cos\theta$. Since $\Sigma = M/A$ (the mass per unit area of the nucleus) varies as $M_o^{1/3}$ the parameter $M_o^{1/3}\mu$ is essentially $\Sigma \cos\theta$ which is its mass per unit area across the comet path projected in the vertical direction in the atmosphere. This is the direction along which we measure the *atmospheric column density* $N$ (or column mass $\Sigma = N\bar{m}$). From Equation (24) for $dM/dN$ — and its equivalent for $dv/dN$ — we see that it is precisely this ratio of atmospheric to projected cometary column masses per unit that occurs in our mass-loss and deceleration equations. This insight also allows an estimation to be made of the effect on our results of nucleus fragmentation which we have neglected above. If the nucleus fragments into say $J$ identical parts, the value of $\Sigma$ for each fragment would be smaller than that of the whole nucleus by a factor $J^{-1/3}$. An upper limit to the effect of this on the termination depth can be set by considering the case when fragments spread apart sufficiently for their bow shocks to be distinct as opposed to all enveloped behind a single shock front. Then the parameter $M_o^{1/3}\mu$ in our equations and Figures for e.g. fragment termination density would be replaced by $J^{-1/3}M_o^{1/3}\mu$, a result which may prove useful in future work on fragmentation.

## 7. Observable Signatures and Information from Sun-Impactors

Having discussed the theory of sun-impacting comet destruction, in this Section we discuss various aspects of their observability. We start here with the likelihood of such impacts in terms of the perihelion frequency distribution $N(q)$ of primary (primordial) comets and of secondary comet streams like Kreutz, then in the following Subsections we discuss some potential observational signatures of impacts in terms of their practical observability and what can be learnt from them. Full quantitative prediction of emission signatures will clearly require detailed numerical simulations like those carried out for the SL-9 Jupiter impacts (eg. Carlson et al. 1995, 1997) and here we confine ourselves to rough estimates to guide future work. One important point is that here we restrict our discussion to observable signatures of impacts which terminate above the photosphere (i.e. at $n < 10^{17}$ cm$^{-3}$) since, even without added ablated material, the atmospheric continuum optical depth exceeds unity below that and the energy release will only be observable indirectly as hot plumes upsurges through, and ripples on, the photosphere. Thus for the purposes of the present section we exclude the parameter regime (cf Equation (27)) $[\mu_{-2}^3 M_{15} X_{-2}] > 1600$ which only applies to rare impacts of unusually high $M_\circ$, $X$, or steep entry angle.

7.1 Comet Perihelion ($q$) Distributions and Likelihood of Impacts

As far as primordial comet perihelia are concerned, both Everhart (1967) and Hughes (2001) report that the observed frequency distribution $N(q)$ differential in $q$ is $q$ independent. Thus, for example the likelihood in the range $0 < q < R_\odot$ is the same as in $R_\odot < q < 2R_\odot$ etc. The perihelia of Group comets (e.g. Kreutz), which have been studied by e.g. Biesecker et al. (2002) Marsden (2005), Lee et al. (2007), Knight et al. (2008, 2010) are quite different, with $q$ values concentrated near $R_\odot$, since they were created by disruption of a larger (primordial?) comet in an earlier near-sun encounter. For example, Lee et al. (2007) find about 85% of SOHO comets to have $q < 2$-$3R_\odot$. Though mostly comprising very small comets the Kreutz group itself includes the Great Comets of 1843 (C/1843 D1) with $q = 1.18R_\odot$ and 1965 (Ikeya-Seki) with $q = 1.67R_\odot$ and Comet Lovejoy 2011 ($q = 1.2R_\odot$). Of course for a group comet to become an impactor it must either be on its first approach to the sun or have had its $q$ value reduced by orbital perturbation — cf discussion by Marsden (1967, 2006) who stated ".. it is certainly possible - indeed probable — that some of them [the Kreutz Group] hit the sun". In summary, solar impactors must exist and should be observable with modern instrumentation, but they are rare and observing them will involve a degree of luck!

## 7.2 Predicting the Time and Place of Impact

We have found the spatial scale of the initial energy release to be ~ $H = 200$ km ($0".3$) vertically and ~ $H/\mu = 20,000/\mu_{-2}$ km ($30"$), while the primary energy release (rise) time is ~ $H/v_o\mu \lesssim 30/\mu_{-2}$ s. *Resolving* the emission from this hot explosion should be easily achievable against the cooler atmospheric background unless or until it is enveloped in opaque material as the energy spreads . Sufficiently accurate absolute pointing and timing to pre-locate the impact site may be harder, depending on how accurately the trajectory is known. Since impactors are necessarily quite large, the incoming Keplerian orbit should be quite accurately known from ground-based data but the trajectory may deviate from that due to non-gravitational effects including anisotropic ablation and fragmentation, (cf. Sekanina & Kracht 2015) and possibly aerodynamic lift.

## 7.3 Atmospheric Response and Radiation Signatures

### 7.3.1 General Points

The energies and masses of comets in the impactor regime are (coincidentally) comparable to the energy releases and masses involved in magnetic flares and Coronal Mass Ejections, the smallest sun-impactors ($M_o$ ~ $10^{12}$ g, $M_o v_o^2/2$ ~ $2\times10^{27}$ erg) corresponding to micro-flares and Hale-Bopp masses ($M_o$ ~ $10^{18}$ g, $M_o v_o^2/2$ ~ $2\times10^{33}$ erg) to super-flares (e.g. Tandberg-Hanssen & Emslie 1988). The 2 keV kinetic energy available per cometary nucleon is ~ 2 keV (thermal equivalent $T$ ~ 20 MK) is also similar to flares. We thus expect some observational similarities between cometary impacts and of magnetic flares though the greater localization, impulsivity and depth in the comet case may result in significant differences. Eichler and Mordecai (2012) have also suggested that the cause of the major 1.2% jump in terrestrial atmospheric $^{14}$C in the year C.E. 775 might have been solar impact of a giant comet of energy > $10^{34}$ erg. Another distinction between cometary impacts and magnetic flares is that impacts can occur at high solar latitudes (Ibadov and Ibodov 2015). We have already summarised in Section 6 our predictions concerning the depth distribution and peak location of energy release gradient as a function of $M_o$, $\mu$ and $X$. Below we discuss more speculatively the response of the atmosphere to and radiation signatures from this input.

7.3.2 Primary and Secondary Atmospheric Responses

Determining the quantitative hydrodynamic and radiative response of the atmosphere to a hypersonic impact clearly demands numerical simulation and we restrict ourselves here to phenomenology and rough estimates (cf. Section 17 of Bronshten 1983 for the case of planetary meteoroids). The primary response, which we consider here, will be fast formation of a localized hot airburst as solar atmospheric gas passes through the bow-sock. Energy from this airburst will propagate outward as prompt electromagnetic radiation (unless or until bottled up by a large increase in optical depth of the surrounding atmosphere as it ionizes), then in a slower secondary phase also involving thermal conduction and mass motion as the expanding hot plume rises. We do not consider this second phase further here as it will require simulations like those by Carlson et al (1995) for the SL-9 Jupiter impacts. In the most likely case of $X$ being small the situation is complicated by the injection of copious ablated high metallicity nucleus material into the post-shock atmospheric gas, enhancing both its radiative opacity and emissivity. We discuss briefly below the possible effects of this complication on observations but first we estimate how high a temperature is likely to be achieved in the primary shock heating.

For very low $C_H$ (large $X$, very little ablation), transfer of energy from the nucleus to the atmosphere is purely via the bow shock, across which (for an *adiabatic* shock in ionized hydrogen) the atmosphere is decelerated from speed $v_o$ to $v_1 = v_o/4$, compressed by a factor of 4, and heated to a shocked temperature given by $kT_{shock} = (3/16)\bar{m}v_o^2/2$ so $T_{shock} \sim 4$ MK. If this $T_{shock}$ is really attained there will be emission of soft X-rays from near the shock and XUV etc from behind that as the gas cools, plus free-free 'radio' emission (THz at these high densities - Fleishman and Kontar 2010). Whether these will actually be observable depends on whether : (a) the shock is really sufficiently adiabatic; (b) there is enough hot gas (emission measure) for the emission to be detectable, if it can escape; (c) the enveloping atmosphere is and remains sufficiently transparent to allow escape of this primary radiation. Question (c) hinges on the larger scale response of the surroundings (especially their opacity) to the primary explosion which we do not attempt to model here. However, we note that in the case of the SL-9 Jupiter impacts, temperatures were observed, at least initially, up to the adiabatic value for $T_{shock} \sim 40,000$ K. To answer (a) and (b) we need to estimate the shock heating timescale $t_{heat}$ and the shocked gas cooling time $t_{cool}$ (dominated by radiation at the high densities involved here). Only if $t_{cool} \gg t_{heat}$ is the shock heating adiabatic enough to attain $T = T_{shock}$. Secondly the value of $t_{cool}$ determines the spatial extent of the hot post-shock region, hence the detectability of its emissions.

The shock heating time $t_{heat} < 4d/v_o$ where shock thickness $d$ is roughly the mean free path $d(cm) \sim 0.1/n_{16}$ with $n_{16} = n(cm^{-3})/10^{16}$ so $t_{heat}(s) < 10^{-8}/n_{16}$. On the other hand $t_{cool}(s) = 3nkT/n^2 f_{rad}(T) > 4 \times 10^{-5} T_6/n_{16} f_{rad-21}$ where $f_{rad}$ (erg cm$^3$ s$^{-1}$) $= 10^{-21} f_{rad-21}$ is the radiative loss function (Pottasch 1965, Cox and Tucker 1969, Cook et al 1989). In the temperature range of interest here $f_{rad}$ has a broad peak dominated by XUV line emission from trace elements like CNO and so is sensitive to the abundance metallicity of the plasma considered. For solar abundances the peak $f_{rad}$ is around $10^{-21}$ erg cm$^3$ s$^{-1}$ so that $t_{cool} > 10^{-4}/n_{16}$ which (regardless of $n$) is about $10^4$ times $t_{heat}$ and there is no doubt that the adiabatic $T_{shock}$ is attained. Behind the shock the heated gas cools on timescale $t_{cool}$ therefore over a distance $s = v_o t_{cool}/4 \sim 10^3/n_{16}$ which is the scale thickness of the post-shock XUV radiating (cooling) region which has a volume $V = sS$ where $S$ is the effective shock area. The total XUV radiative output is $EM f_{rad}$ where "Emission Measure" $EM = n^2 V = n^2 sS$ so, with our estimate of $s$, the post shock XUV source has $EM$ (cm$^{-3}$) $\sim 10^{45} n_{16} S(km^2) \sim 3 \times 10^{45} S(km^2) M_{15}^{1/2} \mu_{-2}^{3/2}$ where we have used Equation (17) for $n_{16}$ at the peak energy deposition depth. Finally if, for the area $S$, we adopt the area of a cylinder of length and radius equal to that of the fully pancaked nucleus (Equation (20)), viz $S(km^2) = 2 \times 10^3 M_{15}^{2/3}$ we get $EM$ (cm$^{-3}$) $\sim 6 \times 10^{48} S(km^2) M_{15}^{7/6} \mu_{-2}^{3/2}$. This is a large XUV (and free-free radio) emission measure by flare standards (Feldman 1996) and the emission should be readily detectable while the shock heating lasts, and as long as it is not masked by enveloping opaque material. If it *is* so masked and if the optical depth unity surface is at a distance *D* then would expect the effective temperature seen to be reduced from $\sim T_{shock}$ to $(S/4\pi D^2)^{1/4}$. Finally we note that, in the high *X* case the emission spectrum from the hot airburst should show little or no enhancement of metallicity above solar since pre-airburst nucleus mass loss is small. However, the decelerated nucleus will sink into the sun and sublimate relatively slowly in the sunlight (like a sun-skimmer), releasing a localized patch of high metallicity deep in the atmosphere.

Turning now to the low *X* case of strong ablation, in addition to the airburst occurring higher (cf Section 6) this differs in several important ways due to the mingling of cooler dense high metallicity ablated matter with the 4 MK post shock solar plasma. This will reduce the airburst temperature due to the heat-sharing with a larger mass and to the large reduction of $t_{cool} \sim kT/nf_{rad}$ by the greatly enhanced *n* value and the major increase in $f_{rad}$ relative to solar by the high metallicity of cometary matter (by a factor $\sim$ 300-500 according to J. Raymond, personal communication). The high metallicity should result in a strong *X*-dependent enhancement of the spectral signatures of the elements involved which might enable a spectral determination of *X* though the interpretation might be complicated by optical depth enhancement by the ablated matter.

## 7.4 Cometary Sunquakes

Isaak (1981) suggested that sun-impactor momentum deposition could generate helioseismic waves and contaminate global helioseismic signals. This idea was discussed briefly by Gough (1994), in the context of

39possible Jovian seismic waves induced by the Shoemaker-Levy 9 impacts, who concluded that only a very large impactor ($\gtrsim 10^{17}$ g ~ Halley) would be detectable in global helioseismic signals of that era. However, that does not preclude detection with modern hardware and analysis methods of transient sun-quake ripples near the sites of lower mass impacts, similar to the *sun-quakes* following some magnetic flares first detected by Kosovichev and Zharkova (1998) and now the extensively studied (e.g. Donea and Lindsey 2005, Lindsey and Donea 2008, Matthews et al. 2011, Lindsey et al. 2014). The relationship between flare and sunquake properties proves to be complex, some small flares producing large quake signatures and vice versa. One theoretical issue is that of radiative damping in the dense chromosphere of downward impulses from explosions at higher levels. The lower altitudes, smaller volumes and faster timescales of cometary airbursts compared to flares, together with their highly directed motion make them seem more likely to generate stronger helioseismic signatures, though this needs to be quantified by detailed modeling.

## Acknowledgments

We thank the anonymous referee for comments improving greatly the content clarity and accuracy of the paper and for suggesting more discussion of the possible observation and value of sun-impactors. The paper has also benefited from discussions with Karl Battams, Paul Bryans, Matthew Knight, John Raymond, Karl Schrijver, and Dimitri Veras.
JCB also gratefully acknowledges the financial support of a Leverhulme Emeritus Fellowship, NASA JPL Visitor Funds and funds from ISSI Bern's

Near Sun Comet Workshop led by Geraint Jones (University College London). He would also like to thank Keri Simpson for valuable proof-reading and comments. Portions of this work were performed at the Jet Propulsion Laboratory, California Institute of Technology, under contract with NASA.

References

Ahrens, T.J., Takata, T., O'Keefe, J.D. Orton, G.S. 1994 Geo RL, 21, 1551A

Alcock, C., Fristrom, C. C., Siegelman, R. 1986, ApJ, 302, 462

Biesecker, D. A., Lamy, P., St. Cyr, O. C., Llebaria, A. & Howard, R. A. 2002, Icarus, 157, 323

Brown, J.C. Potts, H.E., Porter, L.W. & le Chat, G. 2011,A&A 535,71 [B11]

Bronshten, V.A. 1983, *Physics of Meteoric Phenomena*,
    (Geophysics & Astrophysics Monographs, Springer)

Bryans, P. & Pesnell, W.D. 2012, ApJ 760, 18

Carlson, R. W.,Weissman, P. R., Segura, M. et al. 1995, Geophys. Res. Lett. 22, 12, 1557

Carlson R. W., Drossant P., Encrenaz Th. et al. 1997,Icarus, 128,251

Ceplecha, Z., 1966, Bull. Astr. Inst. Czech. 18, 233

Chevalier, R. A. & Sarazin, C. L. 1994, ApJ 429, 863

Chyba, C. F., Thomas, P. J. & Zahnle K. J. 1993, Nature,361, 40

Cook, J. W., Cheng, C.-C., Jacobs, V. L., & Antiochos, S. K. 1989 ApJ 338, 1176

Cox, D.P. & Tucker, W.H. 1969, ApJ 157, 1157

Crawford, D. A. 1996, in: IAU Colloquium 156 — *The Collision of Comet Shoemaker-Levy 9 and Jupiter*, edited by K. S. Noll, H. A. Weaver and P. D. Feldman, pp. 133-156, Cambridge Univ.Press, Cambridge.

Dobrovol'skii, O.V. 1952 Byull, SA0, 6, 11

Donea A.-C. & Lindsey, C., 2005, ApJ 630, 1168

Eichler, D. & Mordecai, D. 2012, ApJ 761, L27

Emslie, A.G. 1978, ApJ 224, 241

Everhart, E. 1967 Astron. J. 72, 1002

Feldman, U., Doschek, G. A., Behring, W. E., & Phillips, K. J. H. 1996 ApJ 460,1034

Field, G.B & Ferrara, A. 1995, ApJ 438, 957

Fleishman, G.D.; Kontar, E. P. 2010, ApJ 709, L127

Gough, D.O. 1994, MNRAS, 269, L17

Howard R.A. et al. 2008 Space Sci. Rev. 136, 67

Hughes, D.W. 2001, MNRAS 326, 515

Ibadov, S. & Ibodov, F.S. 2015, preprint

Isaak, G. 1981, Solar Phys.74, 43

Knight, M.M. 2008, http://www2.lowell.edu/users/knight/docs/dissertation.pdf

Knight, M.M., A'Hearn, M.F.,Biesecker, D.A,, Faury,G. Hamilton, D.P., Lamy, P.,and Llebaria, A. 2010, Astron. J. 139, 926

Kosovichev, A. G., & Zharkova, V. V. 1998, Nature, 393,317

Lee, S. Yi, Y., Kim, Y. H. Brandt, J. C. 2007, JASS 24,227L

Lindsey, C., & Donea, A.-C., 2008, Solar Phys. 251, 627L.

Lindsey, C., Donea, A.-C., Martínez Oliveros, J. C., Hudson, H. S., 2014, Solar Phys. 289, 1457.

McCauley, P.I., Saar, S.H., Raymond, J.C. Ko Y.-K. & Saint- Hilaire, P. 2013, ApJ 768, 161

MacLow, M.-M. & Zahnle, K. 1994, ApJ 434, L33

Marsden, B.G. 1967 Astron. J. 72, 1170 Marsden, B. G., 2005, Ann. Rev. Astron. Astrophys., 43,75

Marsden, B. G. 2006, https://www.ast.cam.ac.uk/~jds/kreutz.htm

Matthews, S. A., Zharkov, S., Zharkova, V. V., Green, L., Pedram, E., 2011, AGU Fall Meeting Abstract #SH51E-02

Nemchinov,I. V. 1967, J. Appl. Math. Mech., 31(2), 320-339

Pesnell W.D. and Bryans P. 2014 ApJ 785, 50

Petrov, C. I. & Stulov, V. P. 1975, Kosmicheskie
    Issledonvaniya, 13, 587

Popova, O.P., Sidneva, S.N., Shuvalov, V.V. & Strelkov, A.S. 2000, Earth, Moon, and Planets, 82-83, 109-128

Pottasch, S.R. 1965, BAN, 18,7

Revelle, D.O. 1979, Journal of Atmospheric and Terrestrial Physics, 41, 453

Richardson, J. E., Melosh, H. J., Lisse, C. M., Carcich, B. 2007, Icarus 191, 176

Schrijver, C., Brown, J.C., Battams, K., et al. 2012, Science 335, 324

Sekanina, Z. 1993, Science 262, 382

Sekanina, Z. 2003, ApJ, 597, 123

Sekanina, Z., & Chodas, P.W. 2012, ApJ 757, 127

Sekanina, Z. & Kracht. R. 2015, ApJ 801, 135

Svetsov, V. V, Nemchinov, I.V., & Teterev, A.V. 1995, Icarus 116, 131

Tandberg-Hanssen, E. Emslie, A.G. 1988, The Physics of Solar Flares Cambridge University Press

Veras, D., Shannon, A., Gänsicke, B. T. 2014, MNRAS, 445, 4175

Weissman, P. R. 1983, Icarus 55, 4

de Winter, D., Grady C., van den Ancker, M.E., Perez, M.R. & Eiroa1, C. 1999, A&A 343, 137

Zahnle, K. 1992: JGR 97, E6, 10

Zahnle, K. & MacLow, M.-M. 1994, Icarus, 108, 1

Figures and Figure Legends

<u>Figure 1</u> Residual fractional mass $M_{final}/M_o$ when velocity $v/v_o \rightarrow 0$ as a function of parameter $X$ for $0.1 < X < 10$. Real $X$ values are probably $<< 0.1$ for which mass-loss is near total before $v/v_o \rightarrow 0$.

<u>Figure 2</u> Relative maximum pancake radius $a/a_o$ and maximum lateral expansion speed $u_{decMaxE'}$ (m/s), of nucleus versus $M_{15}\mu_{-2}^3$, in the analytic large $X$ (deceleration-dominated) regime, at the depth of maximum energy deposition gradient $E'$.

<u>Figure 3</u> Atmospheric density $n_{abl\ maxE'}$ (cm$^{-3}$) in the analytic small $X$ (ablation-dominated) regime at the depth of maximum mass-loss (and energy deposition) gradient versus $M_{15}\mu_{-2}^3$ for (bottom line upward) $X = 0.001$, $0.01$, $0.1$, $1$. Also shown (top line) is the large $X$ (deceleration-dominated) regime result from Equation (17).

<u>Figure 4</u> Heuristic blend (reciprocal mean Equation (30)) of deceleration-dominated and mass-loss-dominated solutions (17) and (27) as an estimate of the end depth $n_{end}(X)$ including both processes.

<u>Figure 5</u> Numerical solutions including both ablation and deceleration for the dimensionless speed $\eta(\xi) = v/v_o$ and fractional mass-loss $\zeta(\xi) = 1-M/M_o$ as functions of scaled dimensionless depth $\xi(N)$ for values of the parameter $X = 0.01, 0.1, 1.0, 10$. These illustrate the change-over from mass-loss dominated to deceleration-dominated as $X$ increases.

Fig 1

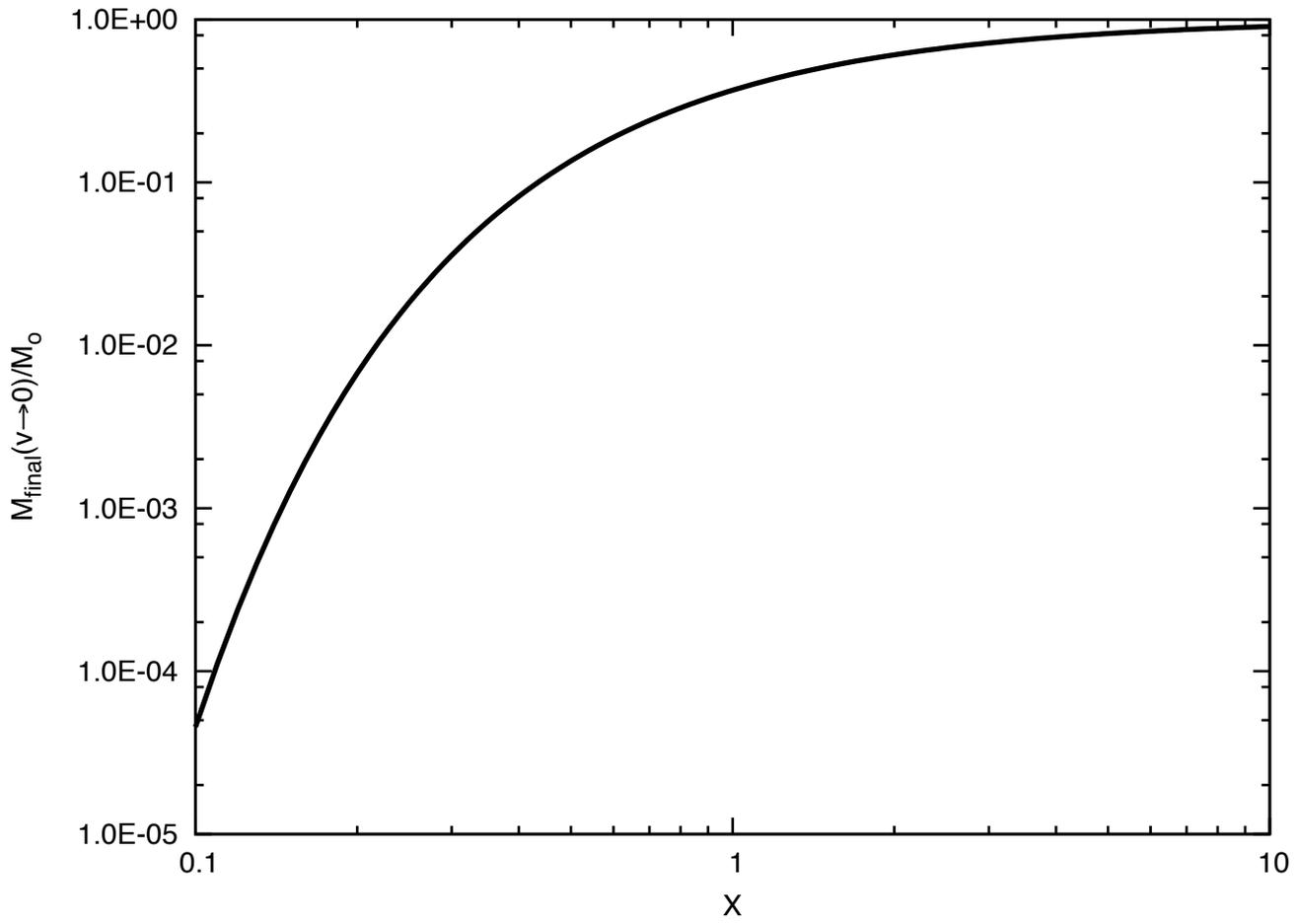

Fig 2

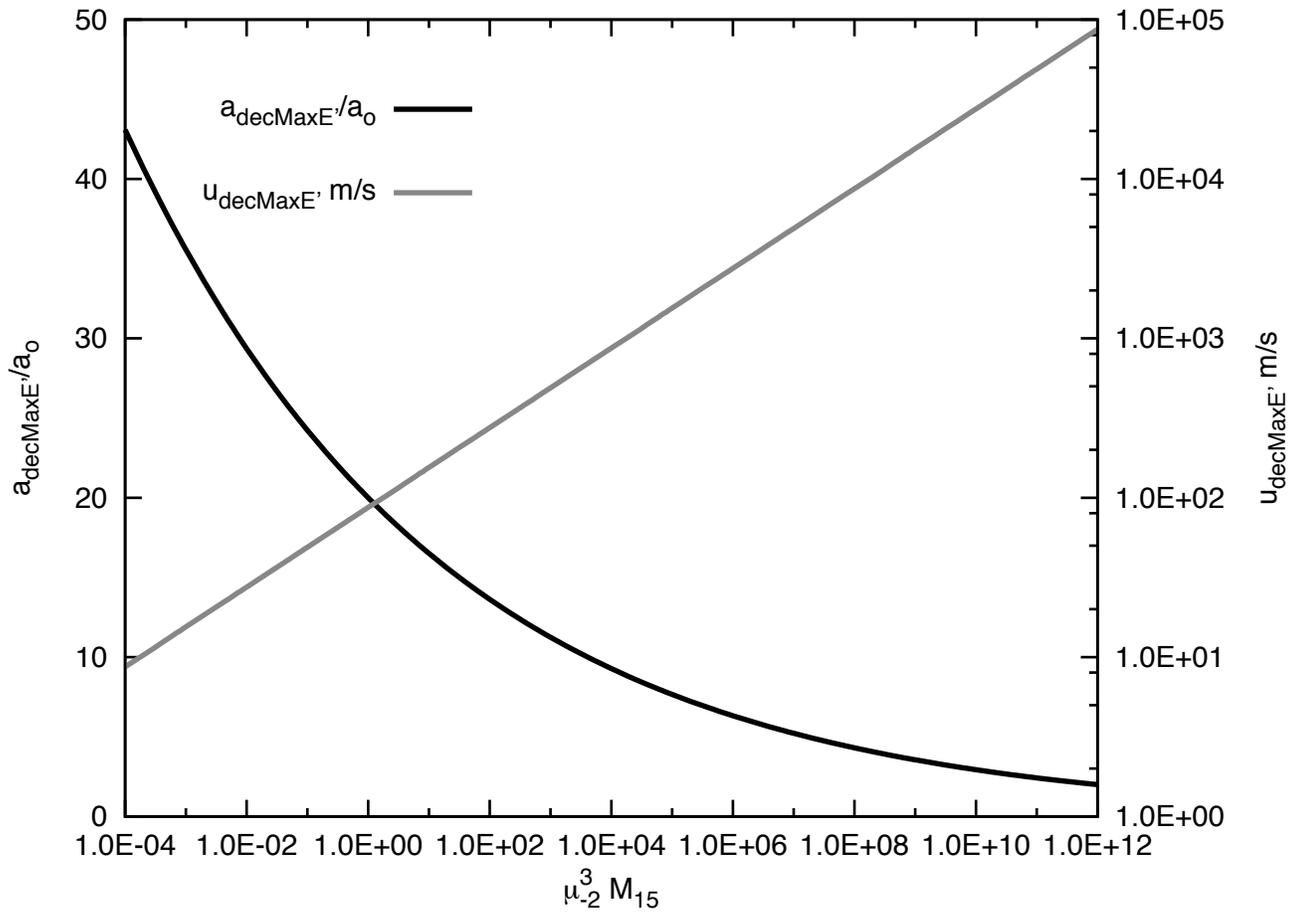

Fig 3

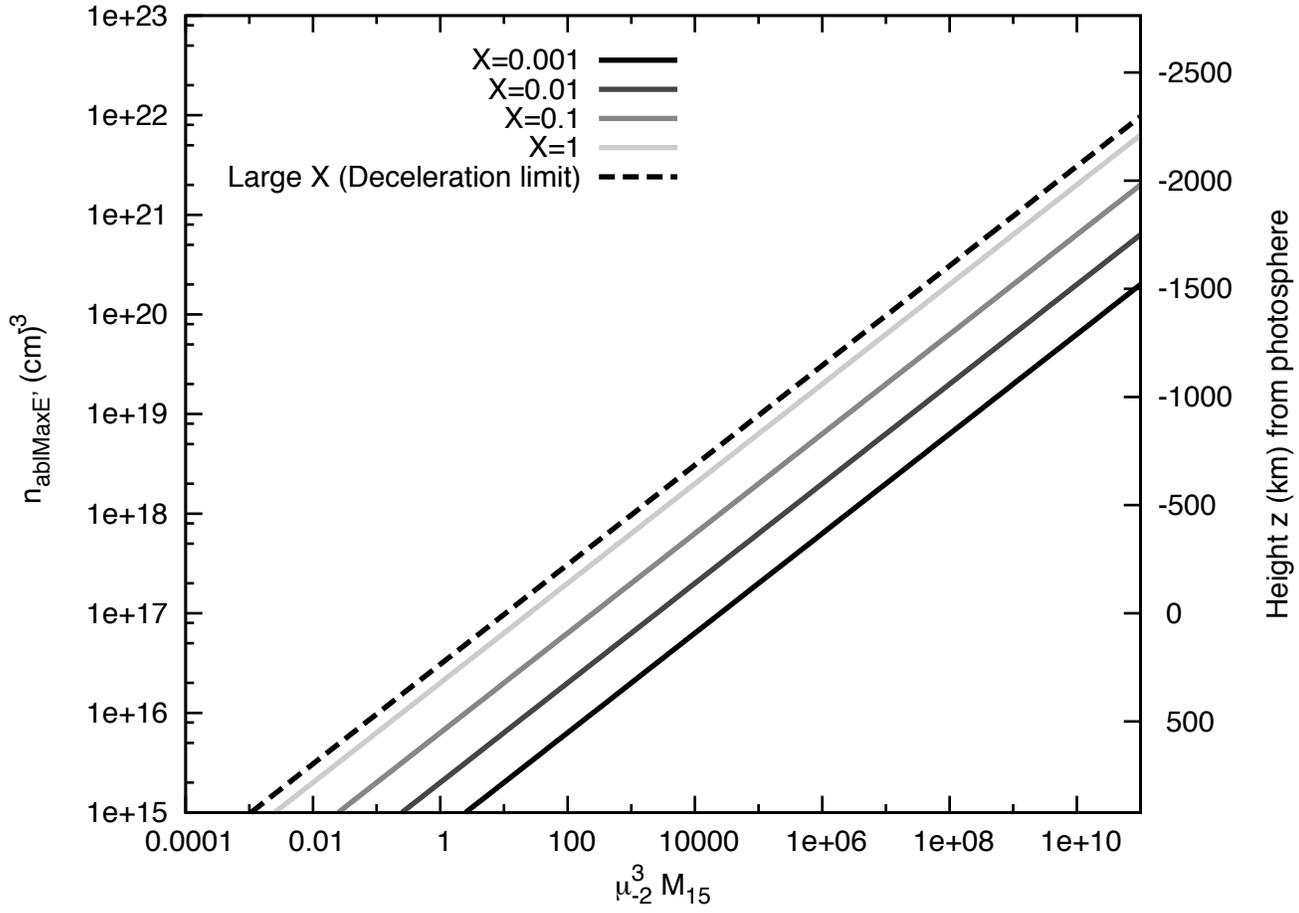

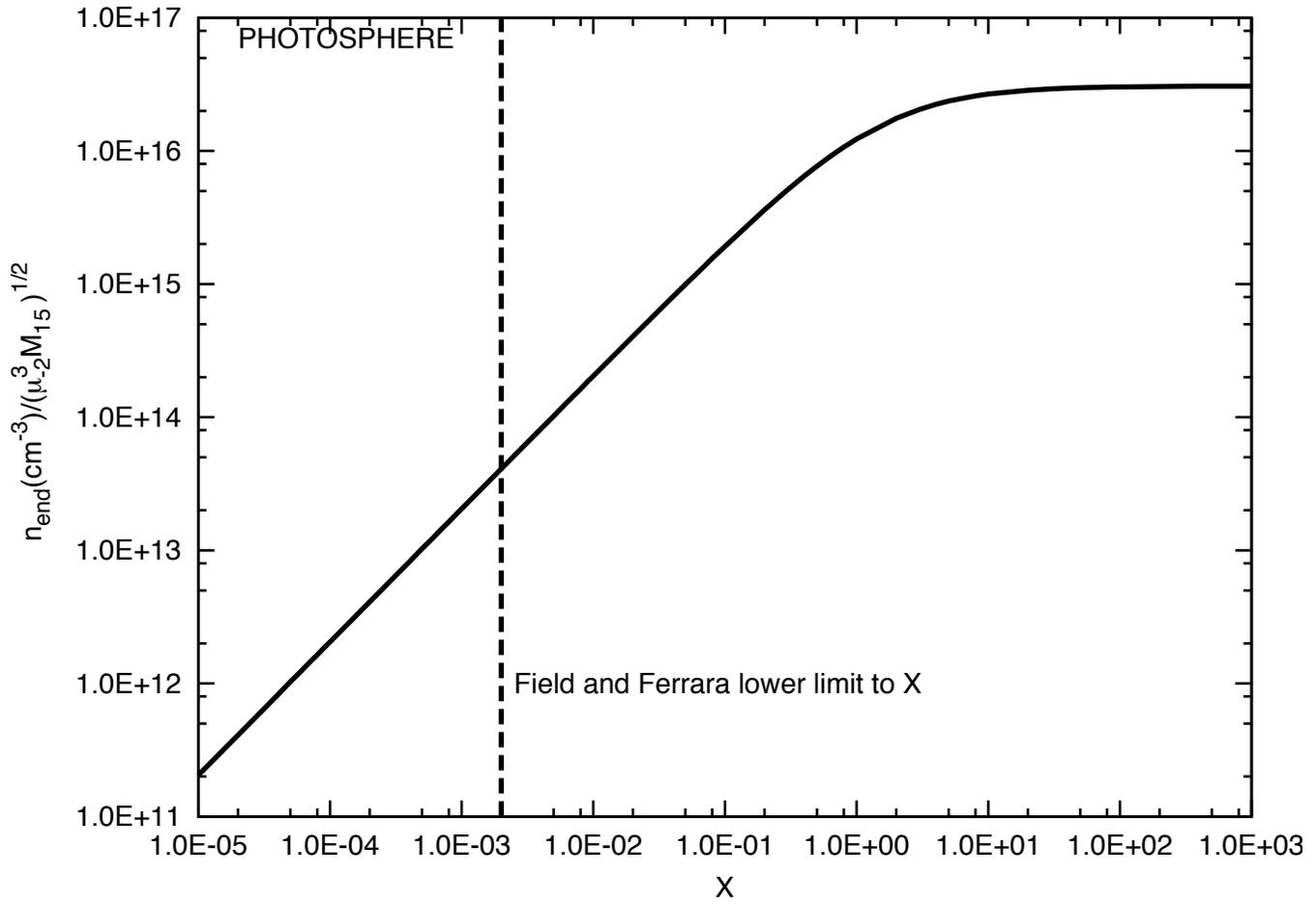

Fig 5

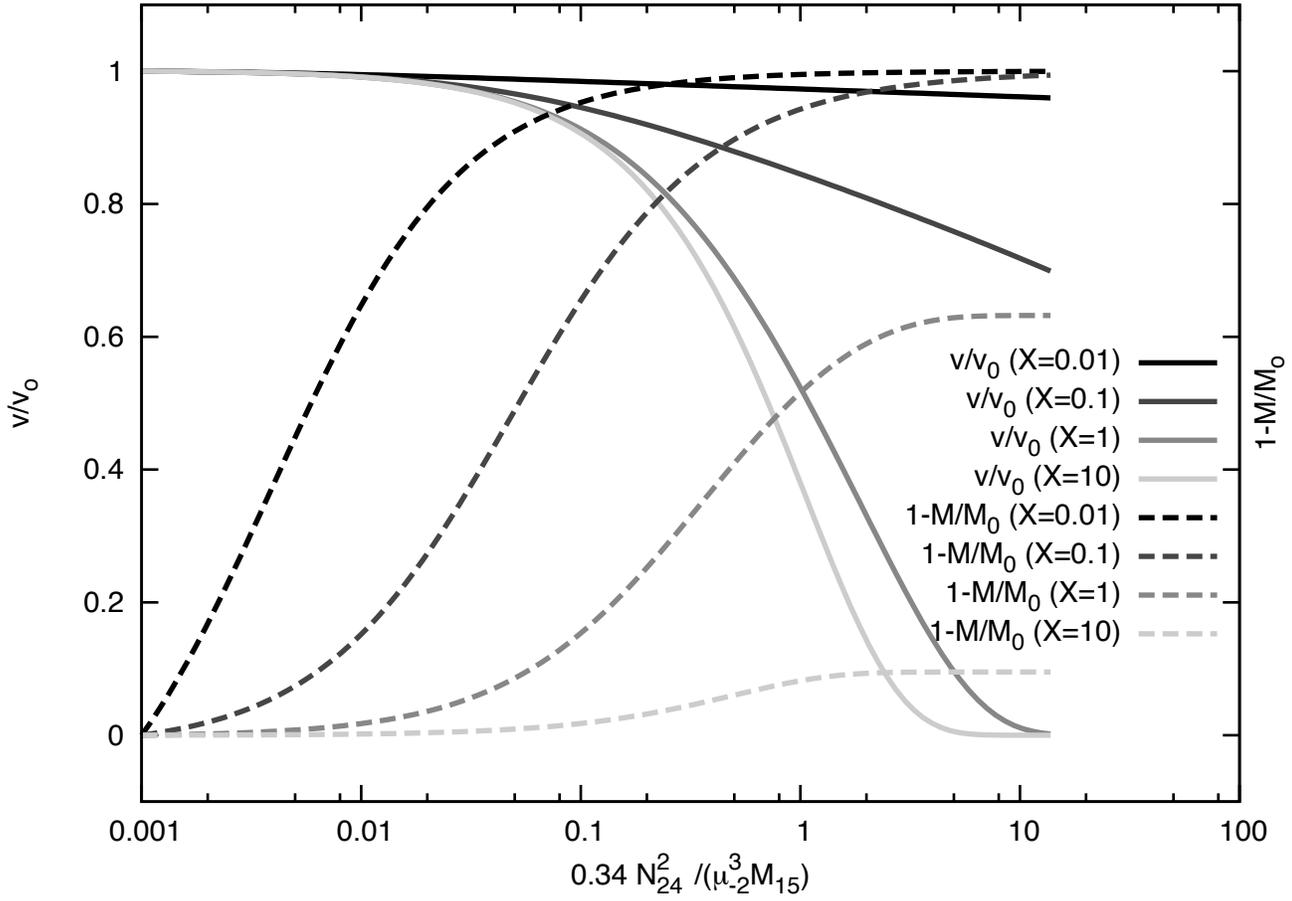